\documentclass[%
 aip,
 amsmath,amssymb,
 reprint,%
]{revtex4-1}

\usepackage{graphicx}
\usepackage{dcolumn}
\usepackage{bm}

\usepackage[utf8]{inputenc}
\usepackage[T1]{fontenc}
\usepackage{mathptmx}

\usepackage{appendix}

\newcommand{\beq}{\begin{equation}}
\newcommand{\eeq}{\end{equation}}
\def\bea{\begin{eqnarray}}
\def\eea{\end{eqnarray}}

\newcommand{\tr}{\text{tr}}

\newcommand{\IR}{{\mathbb R}}

\def\({\left(}
\def\){\right)}

\def\CC{{\cal C}}

\def\CE{{\cal E}}

\def\CG{{\cal G}}
\def\CH{{\cal H}}
\def\CI{{\cal I}}
\def\CJ{{\cal J}}

\def\CN{{\cal N}}
\def\CO{{\cal O}}
\def\CP{{\cal P}}

\def\CR{{\cal R}}

\def\CU{{\cal U}}

\newcommand{\cd}{c^{\dagger}}

\newcommand{\gr}{G^{(\rho)}}
\newcommand{\gs}{G^{(\sigma)}}
\newcommand{\gd}{G^{(\delta)}}
\newcommand{\gns}{G^{(\CN(\sigma))}}
\newcommand{\gnr}{G^{(\CN(\rho))}}

\newcommand{\gfrs}{G^{(\sigma^{\frac 12}\rho\sigma^{\frac 12})}}
\newcommand{\tg}{\tilde \gamma}
\newcommand{\ls}{\lambda^{(\sigma)}}
\newcommand{\lns}{\lambda^{(\CN(\sigma))}}
\newcommand{\lss}{\lambda^{(\sigma^{\frac 12})}}
\newcommand{\pf}{\text{Pf}}

\newcommand{\diag}{\text{diag}}
\newcommand{\tgrs}{\tilde G^{(\sigma \rho)}}

\begin{document}

\preprint{AIP/123-QED}

\title[Recovery Map for Fermionic Gaussian Channels]{Recovery Map for Fermionic Gaussian Channels}

\author{Brian G. Swingle}
 \email{bswingle@umd.edu}

 \affiliation{Condensed Matter Theory Center, Maryland Center for Fundamental Physics, \\
Joint Center for Quantum Information and Computer Science, \\
and Department of Physics, University of Maryland, College Park, MD 20742, USA}
\affiliation{Kavli Institute for Theoretical Physics, Santa Barbara, CA 93106, USA}
\affiliation{Institute for Advanced Studies, Princeton, NJ 08540, USA}

\author{Yixu Wang}%
 \email{wangyixu@terpmail.umd.edu}
\affiliation{ 
Maryland Center for Fundamental Physics,\\ and Department of Physics, University of Maryland, College Park, MD 20742, USA
}%

\date{\today}

\begin{abstract}
A recovery map effectively cancels the action of a quantum operation to a partial or full extent. We study the Petz recovery map in the case where the quantum channel and input states are fermionic and Gaussian. Gaussian states are convenient because they are totally determined by their covariance matrix and because they form a closed set under so-called Gaussian channels. Using a Grassmann representation of fermionic Gaussian maps, we show that the Petz recovery map is also Gaussian and determine it explicitly in terms of the covariance matrix of the reference state and the data of the channel. As a by-product, we obtain a formula for the fidelity between two fermionic Gaussian states. We also discuss subtleties arising from the singularities of the involved matrices.
\end{abstract}

\maketitle

\section{Introduction}

``Operation'' is a common notion in the physical world. Rotating a rigid body or breaking a chinaware plate are examples of classical operations. They can be reversed or not for obvious reasons. On the quantum level, an operation needs to satisfy the complete positivity (CP) condition and the trace preserving (TP) condition to be compatible with the principles of quantum mechanics. Such a compatible operation is called a quantum channel or a CPTP map. Given only this abstract definition, it is less obvious under what conditions we can reverse a given quantum channel. It turns out that for certain channels and input states satisfying an information theoretic criterion, it is possible to fully reverse the channel and recover the initial state. The operation which implements this reversal is called the Petz recovery map~\cite{Petz_1, Petz_2}.

Our motivation for studying recovery channels comes from the recent intensive interaction between quantum information and many other fields of physics, such as quantum field theory and topological quantum matter. The Petz recovery map has important applications in many of these circumstances. For example, one can show that strong subadditivity of the von Neumann entropy implies the average null energy condition (ANEC) in quantum field theory \cite{Faulkner_ANEC}. The Petz recovery map is then related to the conditions for saturation of the ANEC bound. Another example is provided by quantum error correction, say in the context of topological quantum matter, where the Petz map serves as a universal reversal operation that generates no more than twice the error of the optimal reversal operation\cite{BK_near_optimal_reversing}. The Petz map also arises in the structure of quantum Markov states~\cite{QuantumMarkov}, and such Markov states have been used to construct thermal states of quantum many-body systems~\cite{MixedSourcery}. These examples make an explicitly calculable form of the Petz recovery map important in its own right. Here we solve this problem for the special case of Gaussian fermionic states and channels.

The Petz recovery map first originated from considering the notion of sufficiency of channels over von Neumann algebras\cite{Petz_1, Petz_2}. A channel $\CN$ between two algebras $\CN : M\mapsto N$ is sufficient with respect to a family of states $\theta$ if there exists a channel $\beta : N \mapsto M$ such that for any $\phi \in \theta$, $\beta \circ \CN (\phi) =\phi$. $\beta$ is called a recovery channel because it recovers the initial state from the action of the channel $\CN$. It turns out that sufficiency over $\theta$ is equivalent to sufficiency over any pair of the states in $\theta$. And if a channel $\CN$ is sufficient for a pair of states $\{\rho, \sigma\}\in \theta$, then
\beq\label{monoRS}
S(\rho||\sigma)=S(\CN(\rho)||\CN(\sigma)).
\eeq
$S(\rho||\sigma)$ is the relative entropy defined as
\beq
S(\rho||\sigma)=\tr(\rho\log \rho)-\tr(\rho\log\sigma)
\eeq
if $\text{supp}(\rho)\subseteq\text{supp}(\sigma)$, and $+\infty$ otherwise.

For density matrices of finite dimensional Hilbert space and a quantum channel $\CN: M\mapsto N$, the recovery map takes the following explicit form on the support of $\CN (\sigma)$: \cite{HJPW_Structure_SSA}
\beq\label{petz}
\CP_{\sigma,\CN}(X)=\sigma^{\frac 12}\CN^{*}\(\CN(\sigma)^{-\frac 12}X\CN(\sigma)^{-\frac 12}\)\sigma^{\frac 12}.
\eeq
Here $\CN^*: N\to M$ is defined to be the adjoint map of $\CN$ such that
\beq
\langle A, \CN(B)\rangle=\langle\CN^{*}(A), B\rangle
\eeq
for any $A\in N, B\in M$. $\langle A, B\rangle=\tr(A^{\dagger}B)$ is the usual Hilbert-Schmidt inner product. Note that if we restrict the input to $\CP_{\sigma,\CN}$ to be in $\text{supp}\(\CN (\sigma)\)$, then the operator inverse $\CN (\sigma)^{-\frac 12}$ is well defined. Eq.~\eqref{monoRS} is obtained if and only if $\CP_{\sigma,\CN}\circ\CN (\sigma) = \sigma$ and $\CP_{\sigma,\CN}\circ\CN (\rho) =\rho$.

The rotated recovery map was introduced in \cite{wilde2015recoverability}. It is defined as
\beq\label{rotateddef}
\CR_{\sigma,\CN}^{t}(X)\equiv\(\CU_{\sigma,t}\circ\CP_{\sigma, \CN}\circ \CU_{\CN(\sigma),-t}\)(X),
\eeq
where $\CP_{\sigma, \CN}$ is the Petz recovery map and 
\beq
\CU_{\sigma,t}(X)\equiv\sigma^{it}X\sigma^{-it}
\eeq
is a partial isometry group. This rotated recovery map is of interest because it appears in several works that strength the monotonicity of the relative entropy(e.g. \cite{wilde2015recoverability, JRSWW}), and helps to give a refined description of the recoverablity. It is physically interested to investigate as it might give a stronger constraint for energy conditions in quantum field theory\cite{CFQNECtoANEC}.

In this work, we characterize the Petz recovery map $\CP_{\sigma, \CN}$ for fermionic Gaussian channels and states via the Grassmann representation for fermionic Gaussian channels. It turns out that in this case $\CP_{\sigma, \CN}$  is also a Gaussian channel. The same result is established for $\CR_{\sigma,\CN}^{t}$ as well. We provide an explicitly calculable expression which might be useful, for example, for studying error correction in Majorana fermion systems or for studying free fermionic field theories.

Gaussian states are thermal states of Hamiltonians which are quadratic in creation and annihilation operators. More precisely, a state $\rho$ is Gaussian if it has the form $\rho= e^{-\beta H}/\tr(e^{-\beta H})$ for some quadratic $H$ and inverse temperature $\beta$. The ground state of $H$, which is the limiting case as $\beta \to \infty$, is considered a Gaussian states as well. We denote canonical creation and annihilation operators by $\cd,c$. They can be either bosonic or fermionic. The bosonic case could represent, for example, a system of photons, to which the subject of quantum optics is dedicated (e.g.\cite{scully_zubairy_1997}). The fermionic case could describe, for example, electrons or quarks. Here we consider the fermionic case. In either case, Gaussian states have a closure property such that certain so-called Gaussian channels send Gaussian input states to Gaussian output states. The set of Gaussian channels include many common operations such as tracing over subsystems and evolving under quadratic Hamiltonians. Furthermore, Gaussian states can be conveniently prepared in experiments \cite{bachor2004guide} and described by the covariance matrix $G$. A more detailed discussion is presented in Section \eqref{II1}.

To state the main result, let $\gs$ denote the covariance matrix of the reference state $\sigma$. Let a Gaussian channel $\CN$ act on any Gaussian state with covariance matrix $\gr$ such that the output Gaussian state has covariance matrix $\gnr=B \gr B^T +A$, where $A, B$ are matrices encoding the action of channel $\CN$ (see Section \eqref{II2} for details). Our main result is that the corresponding $A, B$ matrices for the recovery channel $\CP_{\sigma, \CN}$ are
\beq\label{PetzAB}
\begin{split}
B_{\CP}&=\sqrt{I_{2n}+(\gs)^2}B^{T}\(\sqrt{I_{2n}+(\gns)^{2}}\)^{-1}\\
A_{\CP}&=\gs-B_{\CP} \gns B_{\CP}^{T}.
\end{split}
\eeq
Based on the above result, the rotated recover map can be characterized by the matrices
\beq\label{RotatedAB}
A_{\CR ,t}=B_{\sigma,t} A_{\CP} B_{\sigma,t} ^T,\quad B_{\CR ,t}=B_{\sigma,t} B_{\CP} B_{\CN(\sigma),-t}
\eeq
in which $B_{\sigma,t}=e^{-2t \arctan \gs}, B_{\CN(\sigma),-t}=e^{2t \arctan \gns}$. It is noteworthy that Eq.~\eqref{PetzAB} is analogous to that for bosonic Gaussian states\cite{LDW_ApproxRev}. 

The rest of the paper is organized as follows: In Section \eqref{II} we give a detailed introduction to Gaussian states and Gaussian channels with fermionic degrees of freedom. In Section \eqref{III}, we construct the Petz recovery map and its rotated version explicitly using the Grassmann representation of the Gaussian map to get Eq.~\eqref{PetzAB},\eqref{RotatedAB}. In Section \eqref{IV}, we present a result giving the fidelity between two fermionic Gaussian states, which is an application of the methods developed in Section \eqref{III}. We draw our conclusions in Section \eqref{V}. We display some identities for Grassmann integrals in Appendix \eqref{A}. Appendix \eqref{B} is a collection of calculation details. In Appendix \eqref{C} we discuss the treatment of singular matrices involved in the derivation.

\section{Fermionic Gaussian states and channels}\label{II}

We first give a brief introduction to the fermionic Gaussian state and fermionic Gaussian channels. Here we mainly follow the results of Ref.~\cite{Bravyi_Lagrangian}. Some other reviews that introduce different aspects of Gaussian states and channels can be found in Refs.~\cite{CEGH_multi_boson, GQI, WHTH_qigs, Greplova_QIFGS}. Interested readers can refer to the references therein.

\subsection{Fermionic Gaussian States}\label{II1}

\subsubsection*{Dirac and Majorana Operators, Grassmann Variables}

We start by considering a Hamiltonian quadratic in complex (Dirac) fermionic creation/ annihilation operators:

\beq\label{quadH}
H=\sum_{i,j}^{n}\cd_i K_{ij}c_j +\cd_i A_{ij}\cd_j +c_i A^{\dagger}_{ij}c_j.
\eeq
Here $\cd$ 's and $c$'s satisfy the canonical anti-communication relation
\beq
\{\cd_{i}, c_{j}\}= \delta_{ij }, \quad   \{c_{i}, c_{j}\}=\{\cd_{i}, \cd_{j}\}=0
\eeq
The matrix $K$ is Hermitian and $A$ is anti-symmetric, so that the Hamiltonian is Hermitian.

It is standard to transform Eq.~\eqref{quadH} into a basis of Majorana fermions:
\beq
\gamma_{2i-1}=(c_i +\cd_i), \gamma_{2i}=i(c_i -\cd_i)
\eeq
which satisfy the anti-commutation relation
\beq \label{MAR}
\{\gamma_i, \gamma_j\}=2 \delta_{ij}.
\eeq
After ignoring the diagonal terms\footnote{The diagonal terms contribute as a number which is the total self-energy of all the modes. It can be absorbed in a redefinition of the zero of energy.}, Eq.~\eqref{quadH} can be written as
\beq\label{Hgamma}
H=\frac i2 \sum_{i,j}^{2n}\gamma_i M_{ij} \gamma_j
\eeq
where
\beq
M=\frac 14 \Im(K) \otimes I_2 + \frac 12 \Re(A)\otimes \sigma_x -\frac i4  \Re(K)\otimes \sigma_y + \frac 12 \Im(A)\otimes \sigma_z .
\eeq
Here $\sigma$'s are the Pauli matrices and $M$ is manifestly real and antisymmetric.

The most general form of the operators composed of $2n$ Majorana operators, denoted as $\CC_{2n}$, should be the complex span of the monomials $\gamma_i \gamma_j ... \gamma_k$
\beq\label{c2n}
X=\alpha \hat I +\sum_{p=1}^{2n}\sum_{1\leq a_1<a_2< ...< a_{p}\leq n} \alpha_{a_1 a_2 ...a_p} \gamma_{a_1}\gamma_{a_2}...\gamma_{a_p}
\eeq
where $\alpha=2^{-n}\tr{X}$ is fixed.  For example, the Hamiltonian in Eq.~\eqref{Hgamma} $H\in \CC_{2n}$, where only the coefficients of order $2$ terms are non-zero and purely imaginary.

To work efficiently with operators built from the Majorana fermions, it is convenient to utilize a Grassman calculus. Grassmann variables are mathematical objects following the anti-commutation rule
\beq\label{GAR}
\theta_i \theta_j +\theta_j \theta_i =0, \quad \theta^2_i =0.
\eeq
For $n$ Grassmann variables $\{\theta_1, \theta_2, ...., \theta_n\}$, denote the complex span of the monomials of these variables as $\CG_n$. A general element $f\in \CG_n$ takes the form
\beq\label{gn}
f=\alpha +\sum_{p=1}^{n}\sum_{1\leq a_1<a_2< ...< a_{p}\leq n} \alpha_{a_1 a_2 ...a_p} \theta_{a_1}\theta_{a_2}...\theta_{a_p}.
\eeq

The similarity between Eq.~\eqref{c2n} and Eq.~\eqref{gn} indicates that there is a natural isomorphism $\omega: \CC_{2n}\to \CG_{2n}$ such that
\beq\label{isoCG}
\gamma_i \gamma_j ... \gamma_k \longmapsto \theta_i \theta_j ...\theta_k, \quad \hat I \longmapsto 1
\eeq
So each Majorana operator $X$ can be mapped via $\omega$ to a polynomial of Grassmann variables $\omega\(X, \theta\)$. We will abbreviate $\omega\(X, \theta\)$ as $X(\theta)$ and call it the Grassmann representation of the operator $X$.

\subsubsection*{Majorana and Grassmann representation of Gaussian states}

Fermionic Gaussian states are in general thermal states for some Hamiltonian of the form in Eq.~\eqref{Hgamma},
\beq
\rho\equiv\tr{\(e^{-\beta H}\)}^{-1} e^{-\beta H}= Z_{\rho}^{-1} \exp \( \frac i2 \beta\sum_{i,j}^{2n} \gamma_i M_{ij} \gamma_j\).
\eeq
This is the Majorana representation of fermionic Gaussian states.

Any real $2n \times 2n$ antisymmetric matrix can be transformed into the following block-diagonal form:
\beq\label{WilliamsonM}
M= O^T\( B\ \otimes \begin{pmatrix}0&-1\\1&0  \end{pmatrix} \)O,
\eeq
where $O\in SO(2n)$ and  $B=\diag(\beta_1,\beta_2,...\beta_n)$. The $\beta_i$'s are the Williamson eigenvalues of the real antisymmetric matrix $M$.

The covariance matrix of a Gaussian state is defined as
\beq\label{covm}
\begin{split}
G_{ij}&=\frac{i}{2}\tr{\(\rho [\gamma_i,\gamma_j]\)}, \text{if $i\neq j$}\\
G_{ij}&=0, \text{if $i=j$}.
\end{split}
\eeq
$G$ is antisymmetric and can be transformed into the block-diagonal form with the same matrix $O$ as in Eq.~\eqref{WilliamsonM}. Its set of Williamson eigenvalues $\{\lambda_1,\lambda_2,...,\lambda_n\}$ are related to $\beta_i$'s via \cite{KWCG_Pairing}
\beq\label{lambdabeta}
\lambda_i =-\tanh \left( \beta \beta_i \right).
\eeq
This induces the matrix equation \footnote{Eq.~\eqref{MtoG} can be found in e.g Ref~ \cite{BGZ_QIgeometry}. We verify a generalization of it in Section~\eqref{matrixrelation} for the completion of the context. A bosonic version of this relation between covariance matrix and the Hamiltonian can be found in Ref.~\cite{BBP15}.}
\beq\label{MtoG}
G=i\tanh \left(i \beta M\right).
\eeq
Since Gaussian states represent the non-interacting limit of the corresponding fermionic degrees of freedom, all higher order correlation functions are totally determined by the matrix element of $G$ by Wick's theorem.

One can use the isomorphism Eq.~\eqref{isoCG} to get the Grassmann representation of the density operator in Eq.~\eqref{Hgamma}. \footnote{The isomorphism Eq.~\eqref{isoCG} can only be used when the operator is written in the form of Eq.~\eqref{c2n}. In this case there are no identical operators in each monomial so the $\gamma_i$'s and $\theta_i$'s are interchangeable.} It turns out that the covariance matrix $G$ plays a role in the Grassmann representation\footnote{One can define displacement operator $D(\mu)$ in terms of Grassmann variables $\mu_i$'s \cite{CG_DOF}. However,the expectation value of observables in the displaced states would involve the product of Grassmann numbers, whose physical meanings are ambiguous to the authors. Though an important completion of the theory, we do not consider it here.}:
\beq\label{Htheta}
\rho(\theta)=\frac{1}{2^n} \exp \(\frac i2 \theta^T G \theta\).
\eeq
The full notational conventions and some useful formula are in Appendix~\eqref{A}.

Eq.~\eqref{Htheta} is a more convenient definition of the Gaussian state, because the matrix $G$ is always bounded by $G^TG\leq I$ according to Eq.~\eqref{lambdabeta}. The Majorana representation Eq.~\eqref{Hgamma} looks singular in the zero temperature limit $\beta \to \infty$, while in the Grassmann representation, this limit corresponds to $\lambda_i \to \pm 1$. This reflects the physical situation that at zero temperature, negative energy modes are occupied and positive energy modes are empty.

\subsection{Fermionic Gaussian Channels}\label{II2}

A linear map $\CN: \CC_{2n}\to \CC_{2n}$ is Gaussian if and only if it admits an integral representation
\beq\label{intrep}
\CN(X)(\theta)=C \int \exp [S(\theta,\eta)+i\eta^T \mu] X(\mu) D\eta D\mu
\eeq
where
\begin{widetext}
\beq\label{channelaction}
S(\theta,\eta)=\frac i2 (\theta^T, \eta^T) N \begin{pmatrix}\theta \\ \eta\end{pmatrix}\equiv\frac i2 (\theta^T, \eta^T)\begin{pmatrix}A&B\\-B^T & D \end{pmatrix}\begin{pmatrix}\theta\\ \eta\end{pmatrix}=\frac i2 \theta^T A\theta +\frac i2 \eta^T D \eta +i \theta^T B \eta.
\eeq
\end{widetext}
Here $A,B,D$ are $2n\times 2n$ complex matrices and $A,D$ can be taken to be antisymmetric. C is a complex number.

To have the linear map be a valid quantum channel, the CPTP conditions should be satisfied. It turns out that the CP condition translates to $C\geq 0$ and requirement that the matrix $N$ in Eq. \eqref{channelaction} be real and satisfy $N^T N\leq I$. The TP condition is equivalent to $C=1$ and the matrix $D=0$. A map is unital if it preserves the identity, which means $\CN(I)=I$. The unital condition is equivalent to $C=1$ and $A=0$.

One important ingredient is how to translate the description of quantum channels from operator representation, e.g., a unitary time evolution written as $e^{i H t} \rho e^{- iHt}$, to the path integral representation defined in Eq.~\eqref{intrep}. \footnote{We draw an analogy of this procedure to transforming from a Hamiltonian formalism to a Lagrangian formalism, as is suggested by the title of Ref.~\cite{Bravyi_Lagrangian}.} This is achieved by making use of the Jamiolkowski duality between linear maps and states.\cite{Jamio_72} The duality says that for any linear map $\CE: \CC_{2n} \to \CC_{2n}$, there is an isomorphism $\CJ$ such that  $\CJ(\CE)\in \CC_{2n}\otimes \CC_{2n}$ takes the form
\beq
\CJ(\CE)=\sum_{i} \CE(V_i)\otimes V_i^{*},
\eeq
where $V_i$'s are a complete set of bases for the linear spaces $\CC_{2n}$, which are the monomials of $\gamma_i \gamma_j ...\gamma_k$.

To get a compact formula, one uses the isomorphism  $\CI: \CC_{2n}\otimes \CC_{2n}\to\CC_{4n}$ such that $\gamma_{p_1}..\gamma_{p_i}\otimes\gamma^{\prime}_{q_1}...\gamma^{\prime}_{q_j}\longmapsto \gamma_{p_1}...\gamma_{p_i}\gamma_{q_1 +2n}...\gamma_{q_j +2n}$, where $1\leq p_1 < ... < p_i\leq 2n$ and $1\leq q_1 < ... < q_j\leq 2n$. This isomorphism induces a map $\CN_1\otimes_f \CN_2: \CC_{4n}\to\CC_{4n}$ such that for monomials
\beq
\begin{split}
\CN_1&\otimes_f \CN_2\(\gamma_{p_1}...\gamma_{p_i}\gamma_{q_1 +2n}...\gamma_{q_j +2n}\)=\\
&\CN_1\(\gamma_{p_1}...\gamma_{p_i}\)\CN_2\(\gamma_{q_1 +2n}...\gamma_{q_j +2n}\).
\end{split}
\eeq
The definition can be extended to polynomials of $\CC_{4n}$ by linearity.

Now let a linear map $\CE: \CC_{2n} \to \CC_{2n}$ be parity preserving, which means it sends even(odd) order monomials to even(odd) order monomials, then the operator $\rho_{\CE}\in \CC_{4n}$ dual to $\CE$ is defined as
\beq\label{operatormapdual}
\rho_{\CE}=\(\CE\otimes_f \hat I\)\(\rho_I\)
\eeq
where
\beq
\rho_{I}=\frac{1}{2^{2n}}\prod_{i=1}^{2n}\(\hat I + i\gamma_i \gamma_{2n+i}\).
\eeq

One can now make use of Eq.\eqref{isoCG} to write Eq.~\eqref{operatormapdual} in the Grassmann representation. The isomorphism is such that $\gamma_1,..., \gamma_{2n} \mapsto \theta_1, ..., \theta_{2n}$ and $\gamma_{2n+1},..., \gamma_{4n} \mapsto \eta_1, ..., \eta_{2n}$. Given the integral representation of the map $\CE$, a straightforward calculation of Eq.~\eqref{channelaction} will give
\beq\label{Grassmannoperatorrep}
\rho_{\CE}(\theta, \eta)=\frac{C}{2^{2n}} \exp \(S(\theta, \eta)\)
\eeq
as the operator representation of a map. Conversely, if one knows the operator representation of the map $\CE$, one can calculate Eq.~\eqref{operatormapdual} explicitly and perform the isomorphism Eq.~\eqref{isoCG} to achieve a form similar to Eq.~\eqref{Grassmannoperatorrep}, so that the integral representation can be read out.

We conclude this section by mentioning an important property of the composition of two Gaussian channels. If $\CE_1$ and $\CE_2$ are CP fermionic Gaussian maps, then the composite map $\CE_2 \circ \CE_1$ is still a CP Gaussian map. Similarly, the composition of two TP Gaussian maps is still a TP Gaussian map. The proof of the results listed in this section can be found in Ref.~\cite{Bravyi_Lagrangian}.

\section{Construction of Petz recovery map}\label{III}
In this section we give the explicit construction of the Petz recovery map, so that the result claimed in the introduction is obtained: the Petz recovery map of a Gaussian channel with a Gaussian reference state is a Gaussian channel itself, and has its information specified in the Grassmann integral representation as in Eq.~\eqref{PetzAB}.

It is obvious from Eq.~\eqref{petz} that $\CP_{\sigma,\CN}$ is composed of three linear maps~\cite{LDW_ApproxRev}: $\CP_{\sigma,\CN}=\CN_3 \circ \CN_2 \circ \CN_1$, where
\beq\label{3maps}
\begin{split}
\CN_1 &:X\mapsto \CN(\sigma)^{-\frac 12} X \CN(\sigma)^{-\frac 12}\\
\CN_2 &: X\mapsto \CN^{*}(X)\\
\CN_3 &: X\mapsto \sigma^{\frac 12}X\sigma^{\frac 12}.
\end{split}
\eeq
The approach to the construction of $\CP_{\sigma,\CN}$ is therefore straightforward. We first find the Grassmann representations of the three separate maps in Eq.~\eqref{3maps}. Since they each admit an integral representation as in Eq.~\eqref{intrep}, the three separate maps are Gaussian linear maps. So it follows that the Petz map is Gaussian. Then we find the formula to combine the three maps together, thus obtaining an explicit expression for the Petz recovery map.

\subsection{Separate construction of three linear maps}

The linear maps $\CN_1$ and $\CN_3$ are of similar form: they represent an operator sandwiched by a Hermitian Gaussian state. Their integral representations are obtained by the map-operator duality described in Section~\eqref{II2}.

We present the construction of $\CN_3$, since that for $\CN_1$ follows in a similar way. The details of the calculation are collected in Appendix~\eqref{cn1n3}.

Consider $\CN_3 : X\mapsto \sigma^{\frac 12}X\sigma^{\frac 12}$ where $\sigma$ is a Gaussian state with covariance matrix $\gs$. Using Eq.~\eqref {WilliamsonM}, define $\tg_i =O_{ij}\gamma_j$ and $\tg_{2n+i} =O_{ij}\gamma_{2n+j}$ where $i,j$ goes from $1$ to $2n$. So the Gaussian state $\sigma$ can be written as
\beq\label{Majrepsigma}
\sigma=\frac{1}{2^n}\prod_{i=1}^{n}\(1-i\ls_i\tg_{2i-1}\tg_{2i}\).
\eeq
One can write $\sigma^{\frac 12}$ as:
\beq\label{sqrtsigma}
\sigma^{\frac 12}=\frac{1}{2^{\frac n2}} \prod_{i=1}^{n}\frac{1}{\sqrt{1+\lambda_{i}^{(\sigma^{\frac 12})2}}}\(1-i\lss_i\tg_{2i-1}\tg_{2i}\).
\eeq
where
\beq\label{sqrtG}
\lss_i=-\lambda_i^{(\sigma)-1}\(\sqrt{1-\lambda_i^{(\sigma)2}}-1\)
\eeq

The operator dual to $\CN_3$ can be explicitly calculated via Eq.~\eqref{operatormapdual} after substitution $\gamma_i \to \theta_i$, $\gamma_{i+2n} \to \eta_i$. Comparing with Eqs.~\eqref{channelaction}, \eqref{Grassmannoperatorrep} and making use of Eq.~\eqref{fg*}, one can read out that the linear map $\CN_3 : X\mapsto \sigma^{\frac 12}X\sigma^{\frac 12}$ corresponds to the following Grassmann integral representation:
\beq \label{n3}
A_3=\gs, B_3=\sqrt{I_{2n}+(\gs)^{2}}, C_3=\frac{1}{2^n}, D_3=-\gs.
\eeq
One can calculate $N_3^T N_3=I$ together with $C_3 >0$ to see that $\CN_3$ is a CP map, where $N_3$ is defined as in Eq. \eqref{channelaction}.

A similar calculation gives the Grassmann integral representation for $\CN_1: X\mapsto \CN(\sigma)^{-\frac 12}X\CN(\sigma)^{-\frac 12}$,
\beq \label{n1}
\begin{split}
 &A_1=-\gns, B_1=\sqrt{I_{2n}+(\gns)^{2}},\\ 
 &C_1=2^n \det\(I_{2n}+\(\gns\)^{2}\)^{-\frac 12}, D_1=\gns.   
\end{split}
\eeq
We have $N_1^T N_1=I$ and $C_1 >0$ so that $\CN_1$ is completely positive as well. Here we have assumed that $\ls \neq \pm 1$ so $\CN(\sigma)^{-\frac 12}$ is invertible. We discuss this in detail in Appendix~\eqref{singularPetz}.

$\CN_2$ is the adjoint map of a given Gaussian linear map $\CN$. As stated in the introduction, $\CN^*$ is defined via $\langle A, \CN(B)\rangle\equiv \langle\CN^{*}(A), B\rangle$, where  $\langle A, B\rangle=\tr(A^{\dagger}B)$. So
\beq
\tr\(X^{\dagger}\CN(Y)\)=\tr\(\CN^{*}(X)^{\dagger}Y\).
\eeq

One can obatian the following Grassmann integral representation of the adjonit map $\CN^{*}$ for a general Gaussian linear map $\CN$:
\beq
A_{\CN^{*}}=D^{\dagger}, B_{\CN^{*}}= B^{\dagger}, C_{\CN^{*}}=C^{\dagger}, D_{\CN^{*}}=A^{\dagger}.
\eeq
For the case that $\CN$ is a quantum channel, i.e. $A, B$ are real, $C=1$ and $D=0$, we get
\beq\label{n2}
A_{2}=0, B_2= B^T, C_2=1, D_{2}=-A
\eeq
The adjoint map of a quantum channel is a completely positive and unital map. This can be explicitly verified by the integral representation of $\CN^{*}$ in Eq.\eqref{n2}. The calculation for this part is in Appendix \eqref{cn2}.

We summarize the integral representation of the three maps:

\begin{widetext}
\beq\label{intrep3maps}
\begin{split}
&A_1=-\gns, B_1=\sqrt{I_{2n}+\(\gns\)^2}, C_1=2^n \det\(I_{2n}+\(\gns\)^2\)^{-\frac 12}, D_1=\gns,\\
&A_2=0, B_2=B^T, C_2=1, D_2=-A,\\
&A_3=\gs, B_3=\sqrt{I_{2n}+\(\gs\)^2}, C_3=\frac{1}{2^n}, D_{3}=-\gs.
\end{split}
\eeq
\end{widetext}

\subsection{Composition of three linear maps}\label{III2}
 Consider two Gaussian linear maps which are specified by $A_1, B_1, C_1, D_1$ and $A_2, B_2, C_2, D_2$. The combination of the two Gaussian linear maps is still a Gaussian linear map. We can readily calculate the corresponding $A_{2\circ 1}, B_{2\circ 1}, C_{2\circ 1}, D_{2\circ 1}$, by performing the $D\theta, D\beta$ integral in the expression

\beq
\begin{split}
  &\CN_2 \circ \CN_1(X)(\alpha)=\int D\beta D\theta D\eta D\mu \times \\
  &\exp\(S_{2}\(\alpha,\beta\)+i\beta^T \theta\) \exp\(S_{1}\(\theta,\eta\)+i\eta^T \mu\)X(\mu)
\end{split}
\eeq

With some algebra, one can find
\beq\label{composite2}
\begin{split}
A_{2\circ 1}&=A_2 +B_2 \(D_2+A_1^{-1}\)^{-1}B_2^T,\\
B_{2\circ 1}&=B_2 \(D_2+A_1^{-1}\)^{-1}A_1^{-1} B_1,\\
C_{2\circ 1}&=C_1 C_2 (-1)^n \pf \(A_1\)\pf \(D_2+A_1^{-1}\),\\
D_{2\circ 1}&=D_1+B_1^T D_2\(D_2+A_1^{-1}\)^{-1}A_1^{-1} B_1.
\end{split}
\eeq
Here we have assumed the invertibility of the matrices $A_1$ and $ \(D_2+A_1^{-1}\)$.

One can make use of Eq.\eqref{composite2} twice to obtain the Gaussian map for the combination of three Gaussian linear maps. We have to do the Grassmann integration four times so we assume the following matrices are invertible: $\gns$, $A+\(\gns\)^{-1}$, $B^T \(A+\(\gns\)^{-1}\)^{-1}B$ and $\(B^T \(A+\(\gns\)^{-1}\)^{-1}B\)^{-1}+\gs$. (Label this set of assumptions as $1^{\prime}-4^{\prime}$.) This is equivalent to assuming the  invertibility of  $\gns$, $A+\(\gns\)^{-1}$, $B$, and $I_{2n}+\(\gns\)^2$. (Label them as $1-4$.) We discuss these assumptions in Appendix \eqref{singularPetz}. One can show that the independent assumptions are $1,3,4$. It turns out that $1,3$ can be overcome by continuity arguments when these matrices are singular. Assumption $4$ is related to the requirement that $\CN(\sigma)$ be invertible. When it is not invertible, we can only determine the Petz recovery map on the support of $\CN(\sigma)$.

After a fair amount of algebra, (collected in Appendix~\eqref{c3maps}.) one can find that the composition of the three maps in Eq.~\eqref{3maps} gives a Gaussian linear map whose integral representation is
\beq\label{intrepPetz}
\begin{split}
B_{\CP}&=\sqrt{I_{2n}+(\gs)^2}B^{T}\(\sqrt{I_{2n}+(\gns)^{2}}\)^{-1},\\
A_{\CP}&=\gs-B_{\CP} \gns B_{\CP}^{T},~ C_{\CP}=1,~ D_{\CP}=0.
\end{split}
\eeq
Note that the form of $A_{\CP}$ guarantees that the reference state $\sigma$ can always be recovered: $\CP_{\sigma, \CN}\circ \CN(\sigma)=\sigma$. It is obvious that the Petz recovery map is trace preserving since $C_{\CP}=1, D_{\CP}=0$. Complete positivity follows from the complete positivity of the three separate maps. So the Petz recovery map is indeed a quantum channel.  The form of Eq.~\eqref{intrepPetz} is quite analogous to that for bosonic Gaussian states obtained in \cite{LDW_ApproxRev}.\footnote{It is plausible to argue that the differences between the formula of the Petz recovery map for bosons and fermions mainly arise from the fact that the covariance matrix for boson is block-diagonalized by $J\in SP(2n)$, while for fermion $O\in SO(2n)$ plays this role, see Eq. \eqref{WilliamsonM}. In this work we do not further our discussion to deriving Eq. \eqref{intrepPetz} based on above group theoretical arguments.}

\subsection{Rotated recovery map}
The definition of the rotated recovery map in Eq.\eqref{rotateddef} explicitly shows that it is a composition of three maps. So we can make use of the techniques developed above to construct the rotated recovery map. 

We first construct the isometry map $\CU_{\sigma,t}$. It is the same method as the construction of $\CN_1$ and $\CN_3$. The first step is the Majorana representation of the operator $\sigma^{it}$.\footnote{In this subsection, we assume that $\sigma$ and $\CN(\sigma)$ are strictly positive operators. This corresponds the constraints of the Williamson eigenvalues $|\ls_i|<1$ and $|\lns_i|<1$ and ambiguities such as $0^{is}$ or the points lying on the radius of convergence are avoided.} It is defined by exponential series:
\beq
\sigma^{it}=\sum_{k=0}^\infty \frac{(it)^k}{k!}\(\log\sigma\)^k
\eeq
in which $\sigma$ is written in the form as in Eq.\eqref{Majrepsigma}. We can further expand $\log \sigma$ since $||i\ls_i \tg_{2i-1}\tg_{2i}||<1$ with $|\ls_i|<1$. After doing the contraction and the re-summation, the result is

\beq
\begin{split}
\sigma^{it}=\frac{1}{2^{int}}&\prod_{i=1}^n \frac 12 \((1-\ls_i)^{it}+(1+\ls_i)^{it}\)\\+&\frac 12 \((1+\ls_i)^{it}-(1-\ls_i)^{it}\)(-i\tg_{2i-1}\tg_{2i}).  \end{split}
\eeq

This agrees with the result of naive analytic continuation from $\sigma^k$, $k\in \mathbb{Z}_+$.

The second step is to perform the calculation that is similar to Eq.\eqref{N3calculation} to read off the Grassmann representation of this isometry. It turns out that
\beq
A_{\sigma, t}= D_{\sigma, t}=0, \quad C_{\sigma, t}=1.
\eeq
In the bases that $\gs$ is block diagonalized, the matrix $B_{\sigma, t}$ is block diagonalized for each mode as well. In each block, the entries are
\begin{widetext}
\beq
B_{\sigma, t}^{block,i}=\begin{pmatrix} \frac 12 \(\(\frac{1-\ls_i}{1+\ls_i}\)^{it}+\(\frac{1-\ls_i}{1+\ls_i}\)^{-it}\)&\frac i2 \(\(\frac{1-\ls_i}{1+\ls_i}\)^{it}-\(\frac{1-\ls_i}{1+\ls_i}\)^{-it}\)\\-\frac i2 \(\(\frac{1-\ls_i}{1+\ls_i}\)^{it}-\(\frac{1-\ls_i}{1+\ls_i}\)^{-it}\)&\frac 12 \(\(\frac{1-\ls_i}{1+\ls_i}\)^{it}+\(\frac{1-\ls_i}{1+\ls_i}\)^{-it}\)\\ \end{pmatrix}
\eeq
\end{widetext}

Making use of the Eq.\eqref{fg*} and noticing that $\log\(\frac{1-x}{1+x}\)=-2 \text{arctanh} x$ for $|x|<1$, we can write $B_{\sigma, t}$ in a basis independent manner as $B_{\sigma, t}=e^{-2t \arctan \gs}$. One can show that $\CU_{\sigma, t}$ is indeed a valid quantum channel.

The last step is to combine three quantum channels in Eq.\eqref{rotateddef} into one, using Eqs.\eqref{composite2} twice. The singular matrices we encounter can be treated by continuity argument as if they were invertible, as is discussed in Section \eqref{III2} and \eqref{singularPetz}. It is straightforward to get
\beq
\begin{split}
 & A_{\CR ,t}=B_{\sigma,t} A_{\CP} B_{\sigma,t} ^T,\quad B_{\CR ,t}=B_{\sigma,t} B_{\CP} B_{\CN(\sigma),-t},\\ 
 & C_{\CR ,t}=1, \quad D_{\CR ,t}=0.
\end{split}
\eeq

\section{Fidelity between two fermionic Gaussian states} \label{IV}
In this section we give a formula for the fidelity of two fermionic Gaussian states in terms of their covariance matrices. This is an immediate application of the quantum map techniques developed above.

Fidelity is a measure of the similarity between two quantum states. We take the definition of the fidelity between $\rho$ and $\sigma$ to be
\beq\label{fidelity}
F(\rho, \sigma)=\tr \(\sqrt{\sigma^{\frac 12}\rho\sigma^{\frac 12}}\).
\eeq
Fidelity is symmetric in the two argument, $F(\rho, \sigma)=F(\sigma,\rho)$, and it is bounded by $0\leq F(\rho, \sigma)\leq 1$~\cite{NielsenChuang}. Two states are identical if $F(\rho, \sigma)=1$ and are orthogonal to each other if $F(\rho, \sigma)=0$.

We can first make use of the map $\CN_3 : X \mapsto \sigma^{\frac 12} X \sigma^{\frac 12}$. For the square root of a Gaussian operator, the construction is the same as that of $\sigma^{\frac 12}$ described in Eq.~\eqref{conssqrt}, and its covariance matrix is obtained by applying Eq.~\eqref{fg*} to Eq.~\eqref{sqrtG}. After a straightforward calculation, we get
\beq\label{fidelityformula}
F(\rho,\sigma)=\frac{1}{2^{\frac n2}}\det(I-\gr\gs)^{\frac 14} \det\(I+\sqrt{I+\(\tgrs\)^{2}}\)^{\frac 14}
\eeq
where $\tgrs\equiv \(\gs+\gr\)\(I_{2n}-\gs\gr\)^{-1}$.

We present the calculation details in Appendix \eqref{cf}. A variation of the formula for the fidelity for fermionic Gaussian states can be found in \cite{BGZ_QIgeometry}.\footnote{The authors do not derive the equivalence of Eq.\eqref{fidelityformula} and the one in \cite{BGZ_QIgeometry}. But a numerical test for the fidelity between random Gaussian states gives the same result for the two formula.} A similar formula for the fidelity of bosonic Gaussian states can be found in Refs.~\cite{BBP15},\cite{PS00}.

\section{Conclusions} \label{V}

In this work, we constructed the Petz recovery map for Gaussian quantum channels with a Gaussian reference state in which the degrees of freedom are fermionic. Using the Lagrangian representation of Gaussian linear maps, we are able to express the Petz recovery map in terms of the covariance matrix of the reference state, and the matrices $A, B$ that encode the information of the Gaussian quantum channel. The main result is collected in Eq.~\eqref{intrepPetz}. As an immediate application of the techniques, we derived a formula for the fidelity of two arbitrary fermionic Gaussian states in terms of their covariance matrices, as is shown in Eq.~\eqref{fidelityformula}. 

The results might be useful for studying error correction in Majorana fermion systems (e.g.~\cite{MajoranaCodes}), or for constructing fermionic recovery maps in experiments. Further directions include applying the recovery map to a quantum field theory system where the degrees of freedom are continuous, and generalizing the formula to approximate Gaussian states as a step towards interacting systems. It would also be interesting to explore the case of approximate recovery maps, on which there has been both theoretical  (e.g.,~\cite{ApproximateRecovery,UniversalRecovery}) and applicational progress (e.g.,~\cite{QSourcechannel}), in the context of fermionic Gaussian states.

\begin{acknowledgments}
We acknowledge a useful exchange with Michael Walter. YW acknowledges Thomas Hartman and Yikun Jiang for comments and discussions during the preparation of this draft. This work is supported in part by the Simons Foundation via the It From Qubit Collaboration.
\end{acknowledgments}

\appendix
\section{Grassmann Calculus Identities}\label{A}
We list some useful formula for Grassmann integral in this section. The integral of a single Grassmann variable is defined as follows:
\beq
\int d\theta \equiv 0, \quad \int \theta d\theta \equiv1
\eeq
For multiple Grassmann variables,
\beq\label{multigrassint}
\int \theta_1 \theta_2 ... \theta_{2n} d \theta_{2n} ... d\theta_2 d\theta _1= 1.
\eeq
The integral is performed from the interior to the exterior.

One usually abbreviates the integral measure $D\theta\equiv d \theta_{2n} ...d \theta_2 d\theta _1$.  Commonly one writes $\theta=(\theta_1, ..., \theta_n)^T$ for multiple variables so
\beq
\theta^T A \eta \equiv \sum_{j,k}^{2n} \theta_j A_{jk}\eta_k .
\eeq

If one makes the following changes of variables
\beq
\eta_i =\sum_{j} R_{ij}\theta _j,
\eeq
the corresponding measure changes as
\beq
D\eta=\det(R)^{-1}D\theta.
\eeq
One can obtain it by requiring that Eq.~\eqref{multigrassint} still holds for the set of variables $\eta_i$'s.

The following formula are useful for the calculation:
\begin{subequations}
\beq \label{traceformula}
\tr(XY)=(-2)^{n}\int D\theta D\mu e^{\theta^T \mu} X(\theta)Y(\mu)
\eeq
\beq
\int D\theta \exp\(\frac i2 \theta^T M \theta \)=i^n \pf(M)
\eeq
\beq\label{grassint}
\begin{split}
\int D\theta &\exp\(\frac i2 \theta^T M \theta +\eta^T B \theta \)\\&=i^n \pf(M) \exp\(-\frac i2 \eta^T  B M^{-1} B^T \eta\)    
\end{split}
\eeq
\end{subequations}

where $\pf$ stands for the Pfaffian. Pfaffian is a polynomial defined for an anti-symmetric matrix $ A$ such that $\pf(A)^2 =\det(A)$. For the odd dimensional case, $\pf(A)\equiv 0$ since $\det(A)=0$. For the even dimensional case, the Pfaffian for a $2n\times 2n$ anti-symmetric matrix $B$ is defined as
\beq
\pf(B)\equiv \frac{1}{2^n n!}\sum_{\sigma\in S_{2n}} \text{sign}(\sigma)\prod_{i=1}^n B_{\sigma_{2i-1},\sigma_{2i}}
\eeq
where $\sigma$ stands for a certain permutation as an element of the permutation group $S_{2n}$, and $\text{sign}(\sigma)$ denotes the parity of the given permutation.

Note that Eq.~\eqref{grassint} holds when $M$ is invertible. However, if $M$ is not invertible, there is a well-defined limit of the right hand side that makes the equality holds. We show this by an explicit calculation.

We work in the bases where the covariance matrix $G$ is in its block diagonal form
\beq
M=\diag\(\lambda_1,..\lambda_p,0,...,0\)\otimes \begin{pmatrix}0&1\\-1&0\end{pmatrix}
\eeq
where the diagonal matrix is $n$ dimensional so it has $(n-p)$ zero Williamson eigenvalues. For simplicity, we take $B=I$, while general $B$ can be transformed into this form by redefining $\tilde \eta =B^T \eta$.

Calculating the Grassmann integral explicitly gives
\beq
\begin{split}
&\int D\theta \exp \(\eta^T \theta+\frac i2 \theta^T M\theta\)\\
=&(-1)^{n}\sum_{k=1}^{p}i^k\sum_{\sigma_1,...,\sigma_k} \(\prod_{i=1}^{k}\lambda_{\sigma_i}\)\\
&\times\eta_{2\bar\sigma_{k+1}-1}\eta_{2\bar\sigma_{k+1}}...\eta_{2\bar\sigma_{p}-1}\eta_{2\bar\sigma_{p}} \(\prod_{i=p+1}^{n}\eta_{2i-1}\eta_{2i}\)
\end{split}
\eeq
where $\sigma_1,...,\sigma_{k}$ label a certain choice of k items from a total number of p and $\bar\sigma_{k+1},...,\bar\sigma_{p}$ are the corresponding left $(p-k)$ items.

Now consider
\beq
M(\epsilon)=\diag\(\lambda_1,..\lambda_p,\epsilon_{p+1},...,\epsilon_n\)\otimes \begin{pmatrix}0&1\\-1&0\end{pmatrix}
\eeq
where all $\epsilon_{p+1},...,\epsilon_n \neq 0$. Now $M(\epsilon)$ is invertible so we can use Eq.~\eqref{grassint} to get
\beq
\begin{split}
&\int D\theta \exp \(\eta^T \theta+\frac i2 \theta^T M(\epsilon)\theta\)\\
&= i^{n}\pf(M(\epsilon)) \exp\(-\frac i2 \eta^T M^{-1}(\epsilon)\eta\)\\
=&(-1)^n i^n \prod_{i=1}^{p}\lambda_i \(1-i\lambda_{i}^{-1}\eta_{2i-1}\eta_{2i}\)\prod_{j=p+1}^{n}\epsilon_{j}\(1-i\epsilon_{j}^{-1}\eta_{2j-1}\eta_{2j}\)\\
=&(-1)^n i^n \prod_{i=1}^{p}\(\lambda_i-i\eta_{2i-1}\eta_{2i}\)\prod_{j=p+1}^{n}\(\epsilon_j-i\eta_{2j-1}\eta_{2j}\)
\end{split}
\eeq

We can now take the limit $\epsilon_{p+1},...,\epsilon_n\to 0$ and it is straight forward to verify
\beq
\begin{split}
&(-1)^n i^n \prod_{i=1}^{p}\(\lambda_i-i\eta_{2i-1}\eta_{2i}\)\prod_{i=p+1}^{n}\(-i\eta_{2i-1}\eta_{2i}\)\\
=&(-1)^{n}\sum_{k=1}^{p}i^k\sum_{\sigma_1,...,\sigma_k} \(\prod_{i=1}^{k}\lambda_{\sigma_i}\)\eta_{2\bar\sigma_{k+1}-1}\eta_{2\bar\sigma_{k+1}}...\eta_{2\bar\sigma_{p}-1}\eta_{2\bar\sigma_{p}}\\&\times \(\prod_{i=p+1}^{n}\eta_{2i-1}\eta_{2i}\)
\end{split}
\eeq
This limit does not depend how $(\epsilon_{p+1},...,\epsilon_n)$ approaches $(0,0, ...,0)$ in a $(n-p)$ dimensional space, so it is well-defined. We have verified that
\beq\label{limiting}
\begin{split}
 \lim_{(\epsilon_{p+1},...,\epsilon_n)\to(0,0, ...,0)} &\int D\theta \exp \(\eta^T \theta+\frac i2 \theta^T M(\epsilon)\theta\)\\
 =&\int D\theta \exp \(\eta^T \theta+\frac i2 \theta^T M\theta\)   
\end{split}
\eeq

So if $M$ is not invertible in Eq.~\eqref{grassint}, we can still write the right hand side formally, which is understood as the unique value obtained by a perturbing and limiting procedure described above and is independent of how we perform the perturbation. Formally, the integration in Eq.~\eqref{grassint} can be viewed as a matrix-valued function of $M$, and the above derivation shows that this function is continuous when $M$ is singular.

\section{Calculation Details}\label{B}

\subsection{Two relations for matrix function} \label{matrixrelation}
We show two matrix function relations whose arguments are anti-symmetric matrices. They are similar to those matrix functions for bosons found in \cite{BBP15}.

Consider an odd function $f: \mathbb{R}\to\mathbb{R}$ and an even function $g: \mathbb{R}\to\mathbb{R}$ whose Talyor series exist at $x=0$. The corresponding function acting on a matrix $M$ is defined via Talyor series. Due to the odd/even property of the function $f$ and $g$, only the odd powers of $f$ and the even powers of $g$ in the Talyor series remain.

Let an anti-symmetric matrix be the argument of the function
\beq\label{X}
X= O^T\( \chi \otimes \begin{pmatrix}0&-1\\1&0  \end{pmatrix} \)O
\eeq
where $O\in SO(2n)$ and $\chi=\diag\(\chi_1,\chi_2,...,\chi_n\)$.

Now define the induced functions $f_*$ and $g_*$ acting on a matrix such that if X takes the form of \eqref{X},
\beq
f_*(X)\equiv O^T\( f(\chi) \otimes \begin{pmatrix}0&-1\\1&0  \end{pmatrix} \)O,~
g_*(X)\equiv O^T\( g(\chi) \otimes I_2 \)O
\eeq
where $f(\chi)=\diag\(f(\chi_1),...,f(\chi_n)\),~g(\chi)=\diag\(g(\chi_1),...,g(\chi_n)\)$.

We want to show
\beq\label{fg*}
f_*(X)=-if(iX),\quad g_*(X)=g(iX)
\eeq

Direct calculation gives
\beq
\begin{split}
&-if(iX)\\
=&\sum_{k=0}^{\infty}\frac{-i}{(2k+1)!}f^{(2k+1)}(0) \(i O^T\( \chi \otimes \begin{pmatrix}0&-1\\1&0  \end{pmatrix} \)O\)^{2k+1}\\
=&\sum_{k=0}^{\infty}\frac{(-1)^{k}}{(2k+1)!}f^{(2k+1)}(0) O^T\(\chi^{2k+1} \otimes \begin{pmatrix}0&-1\\1&0  \end{pmatrix}^{2k+1}\)O\\
=&\sum_{k=0}^{\infty}\frac{1}{(2k+1)!}f^{(2k+1)}(0) O^T\(\chi^{2k+1} \otimes \begin{pmatrix}0&-1\\1&0  \end{pmatrix}\)O\\
=&O^T\(f(\chi)\otimes \begin{pmatrix}0&-1\\1&0  \end{pmatrix}\)O=f_*(X).
\end{split}
\eeq
The same calculation gives the second formula in Eq.\eqref{fg*}.

\subsection{Construction of $\CN_1$, $\CN_3$}\label{cn1n3}

We present the calculation details for the construction of $\CN_3$ in this part, while that for $\CN_1$ follows in a similar way so is largely omitted.

One can write the ansatz for $\sigma^{\frac 12}$:
\beq\label{conssqrt}
\sigma^{\frac 12}=\prod_{i=1}^{n}c_i\(1-i\lss_i\tg_{2i-1}\tg_{2i}\).
\eeq
$c_i$ and $\lss$ are obtained by requiring $\sigma^{\frac 12} \sigma^{\frac 12}=\sigma$. So
\beq
c_i^2=\frac{1}{2\(1+\lambda_{i}^{(\sigma^{\frac 12})2}\)},\quad \frac{\ls}{2}=2 c_i^2 \lss
\eeq
One can solve to get Eq.~\eqref{sqrtsigma} and Eq.~\eqref{sqrtG}.

Note that Eq.~\eqref{sqrtG} has a well defined limit
\beq
\lim_{\ls \to 0}\lss=0
\eeq
It indeed gives the right answer when $\ls =0$. So we will not single out this seemingly singular case in the following derivation.

The operator dual to $\CN_3$ is
\begin{widetext}
\beq\label{N3calculation}
\begin{split}
&\sigma^{\frac 12}\rho_I\sigma^{\frac 12}\otimes_f I\\
=&\frac{1}{2^{3n}}\prod_{i=1}^{n}\frac{1}{1+\(\lss_i\)^2} \(1-i\lss_i\tg_{2i-1}\tg_{2i}\)\(1+i\tg_{2i-1}\tg_{2i-1+2n}\)\(1+i\tg_{2i}\tg_{2i+2n}\)\(1-i\lss_i\tg_{2i-1}\tg_{2i}\)\\
=&\frac{1}{2^{3n}}\prod_{i=1}^{n}1+i\(1-\lambda_{i}^{(\sigma^{\frac 12})2}\)\(1+\lambda_{i}^{(\sigma^{\frac 12})2}\)^{-1}\(\tg_{2i-1}\tg_{2i-1+2n}+\tg_{2i}\tg_{2i+2n}\)\\
-&\(2i\lss_i\)\(1+\lambda_{i}^{(\sigma^{\frac 12})2}\)^{-1}\(\tg_{2i-1}\tg_{2i}-\tg_{2i-1+2n}\tg_{2i+2n}\)+\tg_{2i-1}\tg_{2i}\tg_{2i-1+2n}\tg_{2i+2n}\\
=&\frac{1}{2^{3n}}\prod_{i=1}^{n}1+i\sqrt{1-\lambda_{i}^{(\sigma)2}}\(\tg_{2i-1}\tg_{2i-1+2n}+\tg_{2i}\tg_{2i+2n}\)-i\ls_i\(\tg_{2i-1}\tg_{2i}-\tg_{2i-1+2n}\tg_{2i+2n}\)+\tg_{2i-1}\tg_{2i}\tg_{2i-1+2n}\tg_{2i+2n}
\end{split}
\eeq

Substitute $\gamma_i$ with $\theta_i$, $\gamma_{i+2n}$ with $\eta_i$ and let $\tilde \theta_{2i} \equiv O_{2i,a}\theta_a$, $\tilde \eta_{2i} \equiv O_{2i,a}\eta_a$, etc. One can write

\beq
\begin{split}
&\(\sigma^{\frac 12}\rho_I\sigma^{\frac 12}\)\(\theta,\eta\)\\
&=\frac{1}{2^{3n}}\prod_{i=1}^{n}\(1-i\ls_{i}\tilde\theta_{2i-1}  \tilde \theta_{2i}\)\(1+i\ls_{i}\tilde\eta_{2i-1} \tilde\eta_{2i}\)\(1+i\sqrt{1-\(\ls_i\)^2}\tilde\theta_{2i-1} \tilde\eta_{2i-1}\)\(1+i\sqrt{1-\(\ls_i\)^2}\tilde\theta_{2i} \tilde\eta_{2i}\)
\end{split}
\eeq
\end{widetext}

Comparing with \eqref{channelaction}, \eqref{Grassmannoperatorrep} and making use of \eqref{fg*}, one can read out that the linear map $\CN_3 : X\mapsto \sigma^{\frac 12}X\sigma^{\frac 12}$ corresponds to the  Grassmann integral representation in Eq.~\eqref{n3}.

One can also obtain $\sigma^{-\frac 12}$ by writing down an ansatz and calculating $\sigma^{-\frac 12} \sigma^{\frac 12} =I$. The result is
\beq\label{invsqrtsigma}
\begin{split}
\sigma^{-\frac 12}=2^{\frac n2} \prod_{i=1}^{n}&\frac{1}{\sqrt{1-\(\ls_i\)^2}} \frac{1}{\sqrt{1+\(\lss_i\)^2}}\\
&\times\(1+ i\lss_i\tg_{2i-1}\tg_{2i}\)    
\end{split}
\eeq

After a similar calculation, one can verify that $\CN_1: X\mapsto \CN(\sigma)^{-\frac 12}X\CN(\sigma)^{-\frac 12}$ has the Grassmann integral representation which is specified in Eq.~\eqref{n1}.

\subsection{Construction of $\CN_2$}\label{cn2}
In this part we construct $\CN_2$, which is the adjoint map of a given Gaussian quantum channel $\CN$. We first consider the case when $\CN$ is just a Gaussian map. If $\CN$ is specified by matrices $A,B,D$ and complex number $C$, one can explicitly write
\begin{widetext}
\beq
\begin{split}
\tr(X^{\dagger} \CN(Y))&=C (-2)^{n}\int D\theta D\mu e^{\theta^T \mu} X(\theta) \int \exp [S(\mu,\eta)+i\eta^T \xi] Y(\xi) D\eta D\xi\\
&=C (-2)^{n}\int D\theta D\mu D\eta D\xi \exp[\theta^T \mu +i\eta^T \xi +\frac i2 \mu^T A\mu +\frac i2 \eta^T D \eta +i \mu^T B \eta]X^{\dagger}(\theta)Y(\xi)\\
&=C (-2)^{n}\int D\theta D\mu D\eta D\xi \exp[-i \theta^T \tilde \mu +\tilde \eta^T \xi -\frac i2 \tilde\mu^T A\tilde \mu -\frac i2 \tilde \eta^T D \tilde \eta -i \tilde\mu^T B \tilde\eta]X^{\dagger}(\theta)Y(\xi)\\
&=C (-2)^{n}\int D\theta D\tilde\mu D\tilde\eta D\xi \exp[-i \theta^T \tilde \mu +\tilde \eta^T \xi -\frac i2 \tilde\mu^T A\tilde \mu -\frac i2 \tilde \eta^T D \tilde \eta -i \tilde\mu^T B \tilde\eta]X^{\dagger}(\theta)Y(\xi)\\
&=(-2)^{n}\int D\tilde\eta D\xi e^{\tilde \eta^T \xi}\(C\int\exp[i\tilde \mu^{T}\theta  -\frac i2 \tilde\mu^T A\tilde \mu -\frac i2 \tilde \eta^T D \tilde \eta -i \tilde\mu^T B \tilde\eta]X^{\dagger}(\theta)D\tilde\mu D\theta \)Y(\xi)\\
&\equiv\tr{(\CN^{*}(X)^{\dagger}Y)}
\end{split}
\eeq
\end{widetext}

In the third line we relabel $\tilde\mu_i =i \mu_i$, $\tilde\eta_i =i \eta_i$. The third line goes to the fourth because the measure is changed by a factor $i^{-4n}=1$. Note that  $\exp [S(\mu,\eta)+i\eta^T \xi]$ and integral measures are even in Grassmann numbers, so $X^{\dagger}(\theta)$ is interchangeable with these pieces. However $X^{\dagger}(\theta)$ and $Y(\xi)$ are arbitrary operators, so are not interchangeable.

The Grassmann variables are mapped from Majorana operators which are hermitian, so they should be real. Since $(D\theta)^{\dagger}=d\theta_1 ... d\theta_{2n} =(-1)^{n(2n-1)}D\theta$, the hermitian conjugate of $D\tilde\mu D\theta$ gives an overall factor of $(-1)^{2n(2n-1)}=1$. One gets
\begin{widetext}
\beq
\begin{split}
\CN^{*}(X)(\tilde \eta)&=\(C\int\exp[i\tilde \mu^{T}\theta  -\frac i2 \tilde\mu^T A\tilde \mu -\frac i2 \tilde \eta^T D \tilde \eta -i \tilde\mu^T B \tilde\eta]X^{\dagger}(\theta)D\tilde\mu D\theta \)^{\dagger}\\
&=C^{\dagger}\int\exp[i\tilde \mu^{T}\theta +\frac i2 \tilde\mu^T A^{\dagger}\tilde \mu +\frac i2 \tilde \eta^T D^{\dagger} \tilde \eta +i \tilde\eta^T B^{\dagger} \tilde\mu]X(\theta)D\tilde\mu D\theta .
\end{split}
\eeq
\end{widetext}

Comparing with Eq.~\eqref{intrep} and Eq.~\eqref{channelaction}, we get
\beq
A_{\CN^{*}}=D^{\dagger}, B_{\CN^{*}}= B^{\dagger}, C_{\CN^{*}}=C^{\dagger}, D_{\CN^{*}}=A^{\dagger}.
\eeq
According to Eq.~\eqref{n2}, $\CN^*$ as the adjoint of a quantum channel is obviously unital as $A_2 =0$, $C_2=1$. To see it is completely positive, note
\beq
N_2 =\begin{pmatrix}0& B^T \\ -B& -A \end{pmatrix}=\begin{pmatrix}0& 1 \\ 1& 0 \end{pmatrix} \begin{pmatrix}-A& -B \\ B^T& 0 \end{pmatrix}  \begin{pmatrix}0& 1 \\ 1& 0 \end{pmatrix} =-J N J
\eeq
where $J\equiv\begin{pmatrix}0& 1 \\ 1& 0 \end{pmatrix}$,
\beq
N_2^T N_2 =J N^T N J \leq J^2 =I
\eeq
So the adjont map of a completely positive map is still completely positive.

\subsection{Composition of three maps}\label{c3maps}

Substitute Eq.~\eqref{intrep3maps} into Eq.~\eqref{composite2}, one explicitly gets
\begin{widetext}
\beq
\begin{split}
A_{2\circ 1}&=-B^T \(A+\(\gs\)^{-1}\)^{-1}B,\\
B_{2\circ 1}&=B^T \(A+\(\gns\)^{-1}\)^{-1}\(\gns\)^{-1} \sqrt{I_{2n}+\(\gns\)^2},\\
C_{2\circ 1}&=2^n \det\(I_{2n}+\(\gns\)^2\)^{-\frac 12} (-1)^n \pf \(-\gns\)\pf \(-A-\(\gns\)^{-1}\),\\
D_{2\circ 1}&=\gns-\sqrt{I_{2n}+\(\gns\)^2} A\(A+\(\gns\)^{-1}\)^{-1}\(\gns\)^{-1} \sqrt{I_{2n}+\(\gns\)^2}.
\end{split}
\eeq
The following pieces are useful for further calculations:
\beq
\begin{split}\label{1'4'-4}
-D_3 - A_{2\circ 1}^{-1}&=\(B^T \(A+\(\gns\)^{-1}\)^{-1}B\)^{-1}+\gs\\
&=B^{-1}\(A+\(\gns\)^{-1}+B\gs B^T\) \(B^T\)^{-1}=B^{-1}\(\(\gns\)^{-1}+\gns\)\(B^T\)^{-1},\\
A_{2\circ 1}^{-1} B_{2\circ 1}&=- B^{-1}\(A+\(\gns\)^{-1}\) \(B^T\)^{-1} B^T\(A+\(\gns\)^{-1}\)^{-1}\(\gns\)^{-1} \sqrt{I_{2n}+\(\gns\)^2}\\
&=-B^{-1} \(\gns\)^{-1} \sqrt{I_{2n}+\(\gns\)^2}.
\end{split}
\eeq
One can then explicitly calculate
\begin{subequations}
\begin{eqnarray}
A_{\CP}&=&A_3 +B_3\(D_3 +A_{2\circ 1}^{-1}\)^{-1}B_3^{T}=G-\sqrt{I_{2n}+\(\gs\)^2}B^T \(\(\gns\)^{-1}+\gns\)^{-1} B \sqrt{I_{2n}+\(\gs\)^2},\\
B_{\CP}&=&B_3 \(D_3 +A_{2\circ 1}^{-1}\)^{-1} A_{2\circ 1}^{-1} B_{2\circ 1}=\sqrt{I_{2n}+\(\gs\)^2} B^T  \(\sqrt{I_{2n}+\(\gns\)^2}\)^{-1},\\
C_{\CP}&=&\det\(I_{2n}+\(\gns\)^2\)^{-\frac 12}\pf\(-\gns\)\pf\(-A-\(\gns\)^{-1}\)\nonumber\\
&\times&\pf\(-B^T \(A+\(\gns\)^{-1}\)^{-1}B\)\pf\(\(-B^T \(A+\(\gns\)^{-1}\)^{-1}B\)^{-1}-\gs\). \\ 
D_{\CP}&=&D_{2\circ 1}+B_{2\circ 1}^T D_3 \(D_3 +A_{2\circ 1}^{-1}\)^{-1} A_{2\circ 1}^{-1}B_{2\circ 1}\nonumber\\
&=&\gns-\sqrt{I_{2n}+\(\gns\)^2} A\(A+\(\gns\)^{-1}\)^{-1}\(\gns\)^{-1} \sqrt{I_{2n}+\(\gns\)^2}\nonumber\\
&-&\sqrt{I_{2n}+\(\gns\)^2}\(I_{2n}+A\gns \)^{-1} B \gs B^T \(\sqrt{I_{2n}+\(\gns\)^2}\)^{-1}\nonumber\\
&=&-\(\gns\)^{-1}+ \sqrt{I_{2n}+\(\gns\)^2} \(\gns\)^{-1}
\(A+\(\gns\)^{-1}\)^{-1} \(\gns\)^{-1} \sqrt{I_{2n}+\(\gns\)^2}\nonumber\\
&-& \sqrt{I_{2n}+\(\gns\)^2} \(\gns\)^{-1} \(A+\(\gns\)^{-1}\)^{-1}\(\gns-A\)\(I_{2n}+\(\gns\)^2\)^{-1} \sqrt{I_{2n}+\(\gns\)^2}\nonumber\\
&=&-\(\gns\)^{-1}+\sqrt{I_{2n}+\(\gns\)^2} \(\gns\)^{-1} \(I_{2n}+\(\gns\)^2\)^{-1} \sqrt{I_{2n}+\(\gns\)^2}=0,    
\end{eqnarray}
\end{subequations}
\end{widetext}

$C_{\CP}$ can be calculated by first calculating its square as $\pf\(A\)^2=\det\(A\)$. It's straightforward then that $C_{\CP}^2=1$. We choose $C_{\CP}=1$ because the composition of completely positive maps should still be completely positive. So one can verify that $A_{\CP}$, $B_{\CP}$, $C_{\CP}$ and $D_{\CP}$ are indeed as specified in Eq.~\eqref{intrep3maps}.

\subsection{Fidelity calculation}\label{cf}

We can obtain the Grassmann representation of $\sigma^{\frac 12}\rho \sigma^{\frac 12}$ by direct calculation,
\beq
\begin{split}
&\sigma^{\frac 12}\rho\sigma^{\frac 12}(\theta)\\
=&C \int \exp [S(\theta,\eta)+i\eta^T \mu] \frac{1}{2^n}\exp\(\frac i2 \mu^T \gr \mu\) D\eta D\mu \\
=&\frac{(-1)^{n}}{2^{2n}}\pf(\gr)\pf(G^{(\rho)-1}-\gs)\exp\(\frac i2\theta^T \gfrs \theta\),
\end{split}
\eeq
in which $S,C$ are specified in Eq.~\eqref{n3} and
\beq
\gfrs\equiv \gs +\sqrt{I+G^{(\sigma)2}}\(G^{(\rho)-1}-\gs\)^{-1}\sqrt{I+G^{(\sigma)2}}.
\eeq

Note that
\begin{widetext}
\beq\label{vargfrs}
\begin{split}
&\sqrt{I_{2n}+(\gs)^{2}}\(G^{(\rho)-1}-\gs\)^{-1}\sqrt{I_{2n}+(\gs)^{2}}+\gs\\
=&\sqrt{I_{2n}+(\gs)^{2}}\(\(G^{(\rho)-1}-\gs\)^{-1}+\(\gs+(\gs)^{-1}\)^{-1}\)\sqrt{I_{2n}+(\gs)^{2}}\\
=&\sqrt{I_{2n}+(\gs)^{2}}\(\gs+(\gs)^{-1}\)^{-1}\((\gr)^{-1}+(\gs)^{-1}\)\((\gr)^{-1}-\gs\)^{-1}\sqrt{I_{2n}+(\gs)^{2}}\\
=&\(\sqrt{I_{2n}+(\gs)^{2}}\)^{-1}\(\gs+\gr\)\(I_{2n}-\gs\gr\)^{-1}\sqrt{I_{2n}+(\gs)^{2}}.
\end{split}
\eeq
\end{widetext}

We denote $\delta\equiv \sqrt{\sigma^{\frac 12}\rho \sigma^{\frac 12}}$, so
\beq
\begin{split}
\delta(\theta)=&\(\frac{(-1)^{n}}{2^{2n}}\pf(\gr)\pf(G^{(\rho)-1}-\gs)\)^{\frac 12}\\\times&\prod_{i=1}^{n}\frac{1}{\sqrt{1+\lambda^{(\delta)2}}} \exp\(\frac i2 \theta^T \gd \theta\).
\end{split}
\eeq
By applying Eq.~\eqref{fg*} to Eq.~\eqref{sqrtG}, we get
\beq
\gd=\(\gfrs\)^{-1}\(\sqrt{I+\(\gfrs\)^{2}}-I\).
\eeq

The fidelity of two states $\rho$ and $\sigma$ is merely the trace of the operator $\delta$. Since only the coefficient of $\hat I$ in Eq.~\eqref{c2n} or \eqref{gn} contributes to the trace via $\alpha_X =2^{-n}\tr X$, we focus on that coefficient
\beq
\begin{split}
&\(\frac{(-1)^{n}}{2^{2n}}\pf(\gr)\pf(G^{(\rho)-1}-\gs)\)^{\frac 12}\prod_{i=1}^{n}\frac{1}{\sqrt{1+\lambda^{(\delta)2}}}\\
=&\frac{1}{2^n} \det(I-\gr\gs)^{\frac 14} \det\(I-G^{(\delta)2}\)^{-\frac 14}\\
=&\frac{1}{2^n} \det(I-\gr\gs)^{\frac 14} \det\(\frac 12 \(I+\sqrt{I+\(\gfrs\)^{2}}\)\)^{\frac 14}\\
=&\frac{1}{2^n} \det(I-\gr\gs)^{\frac 14} \det\(\frac 12 \(I+\sqrt{I+\(\tgrs\)^{2}}\)\)^{\frac 14}.
\end{split}
\eeq
The last line makes use of Eq.~\eqref{vargfrs}. Noting that the dimension of $I$ is $2n$, we arrive at the final result in Eq.~\eqref{fidelityformula}.

\section{Treatment of Singular Matrices}\label{C}
As part of the construction of the Petz recovery map and the derivation of the fidelity formula, the inverses of several matrices appear. In the calculations in Appendix~\eqref{B}, we assumed all relevant matrices to be invertible. Here these assumptions are analyzed.

\subsection{Singular matrices in the construction of Petz recovery map}\label{singularPetz}

As discussed in Section.\eqref{III}, we assume 
\beq
\begin{split}
&\gns,~ A+\(\gns\)^{-1},~ B^T \(A+\(\gns\)^{-1}\)^{-1}B,\\ &\(B^T\(A+\(\gns\)^{-1}\)^{-1}B\)^{-1}+\gs    
\end{split}
\eeq
are invertible ($1^{\prime}-4^{\prime}$). We claimed $1^{\prime}-4^{\prime}$ is equivalent to assuming
\beq
 \gns,\quad A+\(\gns\)^{-1},\quad B,\quad I+\(\gns\)^2
\eeq
 are invertible ($1-4$). The equivalence between $1-3$ and $1^{\prime}-3^{\prime}$ is obvious. Eq.~\eqref{1'4'-4} shows $1^{\prime}-4^{\prime}$ implies that $ \(\gns+\(\gns\)^{-1}\) $ is invertible, which is equivalent to $4$ with the help of $1^{\prime}$. One can show $1-4$ implies $4^{\prime}$ just by reversing the derivation in Eq.~\eqref{1'4'-4}. So $1^{\prime}-4^{\prime}$ and $1-4$ are equivalent. We work with $1-4$ hereafter.

We next show that $1,3$ or $1,4$ imply $2$. A neccessary condition for $\CN$ to be completely positive is $A^T A +B B^T \leq I$. If $B$ is invertible, then we have $A^T A < I-B B^T<I$, so $||A||<1$. However, since $||\gns ||\leq 1$, $||\min \lambda^{\(\gns\)^{-1}}||\geq 1$, $A+\(\gns\)^{-1}$ can not have zero eigenvalue so it is invertible. In general $||A||\leq 1$, and  $||\min \lambda^{\(\gns\)^{-1}}||> 1$ when  $I+\(\gns\)^2$ is invertible. So $A+\(\gns\)^{-1}$ is invertible for the same reason. So the independent assumptions are $1,3$ and $4$.

We first assume that $I+\(\gns\)^2$ is invertible. The construction of the Petz recovery map, i.e., given $\gs$, $A$, $B$(or equivalently $\gs$, $B$, $\gns$), find $A_{\CP}$, $B_{\CP}$, $C_{\CP}$ and $D_{\CP}$, can be viewed as a function $\IR^{12n^2}\to \IR^{4n^2}$ for $A_{\CP}$, $B_{\CP}$, $D_{\CP}$ and  $\IR^{12n^2}\to \IR$ for $C_{\CP}$. The matrix multiplication is continuous in its entries just by the continuity of multiplication and adddition. The matrix inverse is known to be continuous when the inverse exists\cite{ContinuityofInverse}.\footnote{By continuity of matrix calculation we mean that each entry of the output matrix when viewed as a multivariable function of the entries of the input matrices is continuous.} So the function $\(\gs,B,\gns\)\to \(A_{\CP}, B_{\CP}, C_{\CP}, D_{\CP}\)$ is continuous when $I+\(\gns\)^2$ is invertible. This continuity guarantees that if we do perturbation $\gns \to \gns (\epsilon_1)$, $B\to B(\epsilon_2)$ so that $\gns (\epsilon_1)$ and $B(\epsilon_2)$ are invertible, then carry on the derivation in Appendix \eqref{c3maps}, and finally take the limit $\epsilon_1 \to 0$ and $\epsilon_2 \to 0$, we will get a well-defined result as in Eq.~\eqref{intrepPetz}.

We then turn to the assumption 4. As is shown in Eq.\eqref{invsqrtsigma}, if 4 does not apply, $\CN(\sigma)^{-\frac 12}$ is not well-defined and $\CN(\sigma)$ can be written as $\CN(\sigma)_{A}\otimes |\psi^{(\CN(\sigma))}\rangle\langle\psi^{(\CN(\sigma))}|_{\bar A}$. Here $A$ is some region in the full Hilbert space $\CH$ while $\bar A$ is its complement. In this case $\CN(\sigma)$ only have support on $A$.  This contradicts with the assumption in \cite{Petz_1, Petz_2} that the state $\CN(\sigma)$ is faithful, which means $\tr\(\CO\CN(\sigma)\)=0$ does not imply $\CO=0$. So usually we assume that $\CN(\sigma)^{-\frac 12}$ is well-defined. If $\CN(\sigma)$ takes the form of $\CN(\sigma)_{A}\otimes |\psi^{(\CN(\sigma))}\rangle\langle\psi^{(\CN(\sigma))}|_{\bar A}$, we can still determine the form of the Petz recovery map on $A$.\cite{LDW_ApproxRev, HJPW_Structure_SSA}

One might worry that if $I+\(\gns\)^2$ approaches a singular limit, then $B_{\CP}$ or $A_{\CP}$ could be unbounded in Eq.~\eqref{intrepPetz}. Actually $B_{\CP}$ will still be bounded, as well as $A_{\CP}$. This can be argued by the complete positive of the composite of completely positivity maps. We give a detailed analysis here. To simplify the discussion, when $I+\(\gns\)^2$ is singular, we take $\CN(\sigma)=|\psi^{(\CN(\sigma))}\rangle\langle\psi^{(\CN(\sigma))}|_{\CH}$ and the discussion can be easily generalized. We first consider the case that $\sigma$ has full support on $\CH$.\footnote{This depends on the choice of the family of states $\theta$ discussed in the introduction. We can always choose $\sigma$ to be the state with the largest support in the family $\theta$.
If $\sigma$ is mixed in some region in Hilbert space $A$, and pure in its complementary $\bar A$, then the considerations below apply for  $A$ and $\bar A$ respectively and $B_{\CP}$ is still bounded.}  We can show that a valid quantum channel that takes such $\sigma$ to $\CN(\sigma)$ can only be a swap, i.e. $B=0$ and $A=\gns$. Consider the monotonicity of the relative entropy. $S(\rho||\sigma)\geq S(\CN(\rho)||\CN(\sigma))$ actually implies that if $\text{supp}(\rho)\subseteq\text{supp}(\sigma)$, then $\text{supp}(\CN(\rho))\subseteq\text{supp}(\CN(\sigma))$. Because otherwise the RHS is $+\infty$ while the LHS is finite so the inequality does not hold. However, since $\CN(\sigma)=|\psi^{(\CN(\sigma))}\rangle\langle\psi^{(\CN(\sigma))}|_{\CH}$ is pure,  for any $\rho$ with $\text{supp}(\rho)\subseteq\text{supp}(\sigma)$, $\CN(\rho)=\CN(\sigma)$. Since $\rho$ is arbitrary, we can conclude that the only quantum channel that does the job is the swap described above. So in the formula for $B_{\CP}$, $B^T$ is strictly a zero matrix so $B_{\CP}=0$ regardless of the singularity of $I+\(\gns\)^2$. If $\sigma$ is a different pure state from $\CN(\sigma)$, then $B$ is an orthogonal matrix with $B B^T =I$. A proper limiting procedure (for example as described in the last part of Appendix \eqref{A}) shows that $B_{\CP}=B^T$, which means that the recovery map rotates the state back. If $\sigma$ and $\CN(\sigma)$ are two identical pure states, we can still show that for each mode $B_{\CP}=B^T, A_{\CP}=0$ or $B_{\CP}=0, A_{\CP}=\gns$. So in any case $B_{\CP}$ is bounded and so is $A_{\CP}$.

\subsection{Singular matrices in the derivation of fidelity formula}
As presented in \eqref{cf}, in the derivation of the fidelity formula, we assume that $\gr$ and $I-\gs\gr$ are invertible. If we make use of Eq.~\eqref{vargfrs}, we might as well assume that $\gs$ and $\sqrt{I+\(\gs\)^2}$ are invertible. However, as discussed in Eq.~\eqref{limiting} and in \eqref{singularPetz}, the continuity property of the Grassmann integral and the final form guarantees that Eq.~\eqref{fidelityformula} is well-defined by a limiting treatment if $\gs$, $\gr$ or $\sqrt{I+\(\gs\)^2}$ are singular. So the physical one is the singularity of the matrix $I-\gs\gr$.\footnote{Equivalently $I-\gr\gs$, as they share the same spectrum.}

We can show that $I-\gs\gr$ being singular implies the fidelity $F(\rho, \sigma)=0$. One can make use of the trace formula Eq.~\eqref{traceformula} to calculate
\beq
\begin{split}
&\tr (\rho \sigma)\\=&\frac{(-2)^{n}}{2^{2n}}\int D\theta D\mu  e^{\theta^T \mu} \exp\(\frac i2 \theta^T \gr\theta\)\exp \(\frac i2\mu^T \gs \mu\)\\
=&\frac{1}{2^n}\pf(\gr)\pf(G^{(\rho)-1}-\gs)=\frac{1}{2^n}\det\(I-\gr\gs\)^{\frac 12}.
\end{split}
\eeq
So when $I-\gs\gr$ is singular, we have $\tr (\rho \sigma)=0$. However $\tr (\rho \sigma)=\tr(\sigma^{\frac 12}\rho \sigma^{\frac 12})$ and $\sigma^{\frac 12}\rho \sigma^{\frac 12}$ is a positive semidefinite operator so all its eigenvalues $\lambda_i^{(\sigma^{\frac 12}\rho \sigma^{\frac 12})}\geq 0$. So $\tr(\sigma^{\frac 12}\rho \sigma^{\frac 12})=0$ implies that $\sigma^{\frac 12}\rho \sigma^{\frac 12}=0$ since the only case is that all its eigenvalues are $0$.  Then naturally its square root is $0$ so the fidelity $F(\rho, \sigma)=0$.

\bibliography{Petz_bib}

\end{document}